# UV-induced mutagenesis in *Escherichia coli* SOS response: A quantitative model

Sandeep Krishna [1], Sergei Maslov [2], Kim Sneppen [1]

[1] Niels Bohr Institute,
University of Copenhagen,
Blegdamsvej 17, 2100 Copenhagen Ø, Denmark.

[2] Department of Condensed Matter Physics
and Material Sciences,
Brookhaven National Laboratory,
Upton, New York 11973, USA

## Summary

*Escherichia coli* bacteria respond to DNA damage by a highly orchestrated series of events known as the SOS response, regulated by transcription factors, protein-protein binding and active protein degradation. We present a dynamical model of the UV-induced SOS response, incorporating mutagenesis by the error-prone polymerase, Pol V. In our model, mutagenesis depends on a combination of two key processes: damage counting by the replication forks and a long term memory associated with the accumulation of UmuD'. Together, these provide a tight regulation of mutagenesis resulting, we show, in a "digital" turn-on and turn-off of Pol V. Our model provides a compact view of the topology and design of the SOS network, pinpointing the specific functional role of each of the regulatory processes. In particular, we explain the recently observed second peak in the activity of promoters in the SOS regulon (Friedman *et al.*, 2005, PLoS Biol. **3**, e238) as the result of a positive feedback from Pol V to RecA filaments.

## Synopsis

Ultraviolet light damages the DNA of cells which prevents its duplication and thereby cell division. Bacteria respond to such damage by producing a number of proteins that help to detect, bypass and repair the damage. This SOS response system displays intricate dynamical behavior, in particular the tightly regulated turn-on and turn-off of error-prone polymerases which result in mutagenesis, and the puzzling resurgence of SOS gene activity 30-40 minutes after irradiation. In this paper, we construct a mathematical model that systematizes the known structure of the SOS subnetwork based on experimental facts, while remaining simple enough to illuminate the specific functional role of each regulatory process. We can thereby identify the protein-protein interactions and the positive feedback mechanism that are particularly important for the on-off nature of mutagenesis.

## Introduction

The SOS response in the bacterium *Escherichia coli* encompasses many proteins involved in detecting and repairing DNA damaged by a variety of agents, such as UV radiation, or chemicals like mitomycin and bleomycin [1]. A complex regulatory network, comprising both transcriptional and post-translational regulators, controls the concentrations and levels of activity of these proteins (Fig. 1.)

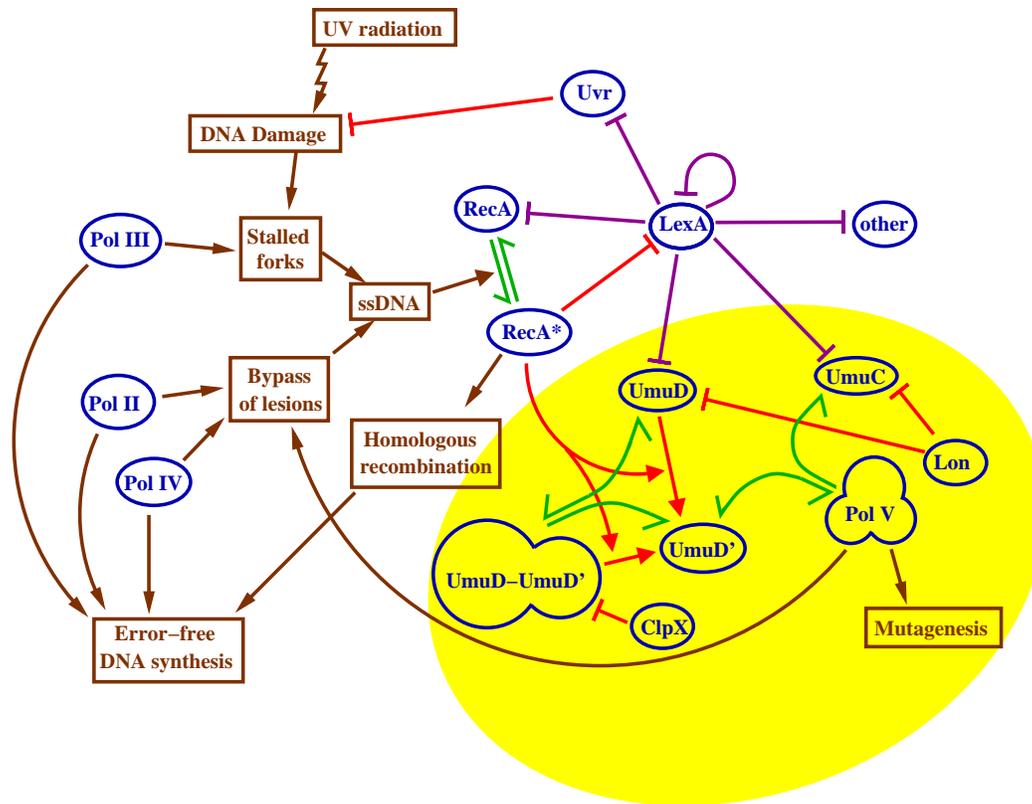

**Figure 1. Schematic representation of the SOS network in *E. coli*, including proteins, functional states of DNA, and key processes. The purple lines indicate transcriptional regulation, the red lines – active degradation and proteolytic cleavage, and the green lines – complex formation. The yellow shading highlights the proteins involved in mutagenesis, centered around the Pol V DNA polymerase, a complex consisting of an UmuD' homodimer and UmuC.**

The collective actions of this regulatory network are orchestrated so that the SOS response is commensurate to the magnitude of DNA damage [1]. Mutagenesis, such as the introduction of single-base substitutions in the DNA sequence, is not an inevitable consequence of DNA damage but, rather, results from the action of specialized error-prone DNA polymerases that are part of the response [2]. This constitutes an extreme measure that might be useful for the cell only after very heavy DNA damage when DNA replication and repair cannot effectively proceed without it. While some mutations might benefit the offspring, the vast majority are harmful, therefore the presence of error-prone polymerases should be tightly regulated to prevent their action at low doses of UV.

Briefly, the sequence of events triggered by UV irradiation of *E. coli* is as follows: UV radiation damages the DNA by creating lesions that mechanically disrupt the process of DNA duplication by stalling the DNA-polymerase (Pol III) in a moving replication fork. This, in turn, results in the production of single-stranded DNA (ssDNA) gaps. These gaps are coated by the protein RecA [1, 3, 4], forming long nucleoprotein filaments in which it assumes its active form RecA*. RecA*, together with other proteins, is involved in the non-mutagenic filling in of ssDNA gaps via homologous recombination [5], and it catalyses the cleavage of the transcriptional repressor LexA [6] and of the protein UmuD [7], whose cleaved form – UmuD' – is necessary for mutagenesis [1]. The drop in the level of the transcription factor LexA, due to its cleavage, de-represses the regulon involved in the SOS response. This regulon comprises around 30 genes, including those encoding the mutagenesis proteins UmuD and UmuC, RecA, and LexA itself. Also part of the SOS regulon are genes encoding UvrA, B, C – a group of Nucleotide Excision Repair (NER) proteins that locate and excise damaged regions from the DNA [8][9].

Mutagenesis in UV-irradiated *E. coli* cells is mainly the direct result of the activity of the error-prone DNA polymerase Pol V [2]. Pol V consists of two units of UmuD' (UmuD protein cleaved by RecA$^*$) and one unit of UmuC. It inserts several random base pairs in the DNA strand directly opposite to a lesion, thus helping a replication fork to quickly bypass the lesion after which Pol III can take over and continue replication. A distinct coordinated sub-network of proteins centered on UmuD and UmuC control the abundance, and thereby, the activity of Pol V (Fig. 1).

Even though the SOS response in bacteria has been studied for several decades, new discoveries continue to be made. Recent single-cell experiments measured the temporal dependence of the activity of LexA-regulated promoters [10] which showed the following features: For low UV doses, the promoter activity peaks at around 10 minutes after the UV dose. This was also observed in bulk measurements that average promoter activity over a large population of cells [4] and can be attributed to the initial rapid drop in LexA levels after UV damage because of the activation of RecA, followed by a slow increase to its original level as the lesions are repaired by NER and the level of RecA* falls. More surprising was the observation that at higher doses of radiation, LexA-regulated promoter activity often had a second peak at around 30-40 minutes sometimes even followed by a third peak at 60-90 minutes. This resurgence of the SOS response is puzzling because it indicates a temporary increase in RecA* levels at a time when the NER process is well under way and the number of lesions is already falling. This second peak (but not the third peak) was, however, absent in both $\Delta$UmuDC null-mutants as well as mutants which have an uncleavable version of UmuD (K97A) [10]. The common element in both types of mutants is the absence of Pol V, which suggests that the second peak is related to mutagenesis.

In this paper we propose a plausible mechanism for the appearance of this peak. We argue that *E. coli* bacteria can reliably measure the total amount of DNA damage. The ability of replication forks to bypass bulky lesions allows the cell to "count" the number of lesions they encountered over a fixed time interval (the average lifetime of RecA* filaments). The result of this count, given by the instantaneous number of RecA* filaments, is then fed into the mutagenesis regulatory sub-network, which, as we show below, is designed to time-integrate this input signal over a long interval (30-40 minutes) and to abruptly turn on the Pol V if the integrated level of damage exceeds some critical threshold. The appearance of Pol V speeds up the bypass of lesions, and thus increases the rate at which new lesions are encountered by replication forks. We believe that this positive feedback from Pol V to the RecA* concentration is responsible for a temporary increase in the activity of SOS-regulated promoters 30-40 minutes after the radiation (the second peak reported in Ref. [10].)

## Approach and Results

### *Structure of the model*

The goal of this paper is to model temporal dynamics of the mutagenesis sub-network of the SOS response system (highlighted in yellow in Fig. 1) for different doses and durations of UV radiation. This sub-network is not isolated from the rest of SOS response, and therefore the model also includes other parts of the entire *E. coli* regulatory network that interact with proteins involved in mutagenesis. Figure 1 shows the components of the SOS response that we quantify in our model. Different colored arrows correspond to different mechanisms of interactions between the nodes. An excellent earlier paper by Aksenov [11] contains a model of LexA-controlled transcriptional regulation coupled with the NER repair of lesions during the SOS response. Here that model is extended to incorporate the mutagenesis sub-network. Full details of our model and parameter values are provided in the Methods section.

We mathematically model the temporal dynamics of the density of UV-induced lesions, as well as concentrations of LexA, RecA*, unbound UmuD, unbound UmuD', UmuD-UmuD' heterodimer, and Pol V, using a set of ordinary differential equations. Positive and negative terms in these equations represent different ways of production and consumption/degradation of the corresponding quantities. We do not explicitly simulate the creation and repair of individual lesions, nor do we simulate each replication fork moving along the DNA. Thus, our model ignores stochastic fluctuations. However, in later sections we do examine the effect of averaging over a population of cells in which various parameters, e.g. the number of replication forks, vary from cell to cell. This provides an *in silico* comparison between single cell and cell culture measurements. We also treat all time-delays, such as when a replication fork is stalled at a lesion, in a simplified manner, i.e., we assume that these delays affect the RecA* level only via the average replication speed.

Most parameters in our model have been fixed using experimental data. For example, the experiments in refs. [3, 4, 12] allow us to fix the RecA*-mediated cleavage rates of LexA and UmuD. The model has a total of 18 parameters of which only 3 could not be fixed by experimental data. We have therefore scanned a range of reasonable values for these three as described in a later section.

Our model indicates four key features of the mutagenesis sub-network in *E. coli*:

1. A mechanism for measuring the local amount of damage, coupling the number of RecA* filaments to the current lesion density.
2. A long term "memory" used to time-integrate the RecA* signal and thus to determine whether damage level remained high for a substantial time. This mechanism is based on slow accumulation of UmuD'.
3. Strong binding between UmuD and UmuD', which provides a highly ultrasensitive increase in unbound UmuD' levels as its concentration exceeds that of its "inhibitor" UmuD.
4. Positive feedback from Pol V to RecA* levels, which further increases the sharpness of the turn-on and turn-off of Pol V. This mechanism is also responsible for the second peak in activity of SOS promoters.

In the subsequent sections we discuss each of the above aspects in more detail.

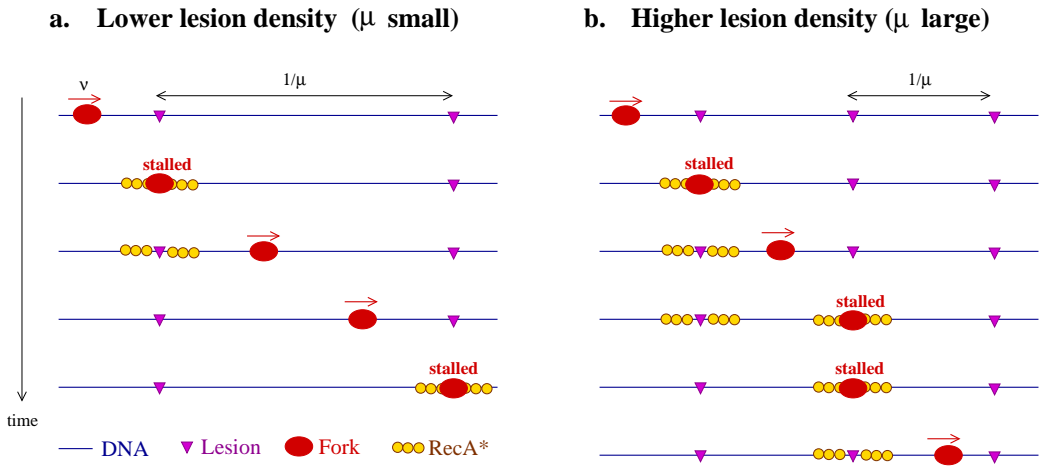

**Figure 2. Mechanism for measuring damage.** When the replication fork stalls at a lesion, a RecA filament (RecA*) is formed. Each filament exists for an average time $\tau_{RecA*}$. For low lesion densities (a) the filament disassembles before the replication fork reaches the next lesion. In contrast, an extreme scenario is depicted in (b) where the lesion density is so high that the replication fork reaches the next lesion before the first filament disassembles. In this case more than one RecA filament can be present on the DNA for some time, and the average RecA* concentration is correspondingly higher. For intermediate lesion densities, the average concentration of RecA* also increases with the lesion density, its value being determined by the interplay between the stall time ($\tau_{stalled}$), distance between lesions ($1/\mu$), speed of the fork on undamaged DNA ($v$) and the RecA filament lifetime ($\tau_{RecA*}$) (see main text).

## *Measuring damage*

First we propose the following mechanism for the influence of the UV dose on the RecA* level: Consider a given replication fork proceeding on a DNA strand which has UV-induced lesions, as depicted in Figure 2. The Pol III DNA-polymerase stalls at the first lesion thus generating an ssDNA gap, which is then covered with RecA. This RecA filament exists for an average time, denoted $\tau_{RecA*}$, after which it disassembles (we assume that each filament disassembles independently with a rate that is not limited by other DNA damage induced processes). During this time the replication fork may bypass the lesion and continue processing the DNA, leaving the first RecA filament behind. If the time the fork spends stalled at a lesion is sufficiently large or the lesion density is sufficiently small (so that the time the fork spends traveling between lesions is large), then the first filament will disassemble before the fork reaches the next lesion and creates another filament (as in Fig. 2a). Therefore, in this case, there will only be no more than one RecA* filament per replication fork at any time. On the other hand, if the stall time is small or the lesion density is large, the fork will reach a second lesion before the first filament disassembles and, as a consequence, there may be many RecA* filaments per fork existing simultaneously on the DNA (as in Fig. 2b).

The RecA* level directly depends on the time a polymerase spends traveling between lesions, $\tau_{moving} = 1/\mu v$, where $\mu$ is the density of lesions on the chromosome, and $v$ is the average speed with which Pol III processes DNA replication on undamaged DNA. This dependence can quantified: One RecA* filament is produced every time the replication fork encounters a lesion. If the fork spends a time $\tau_{stalled}$ at a lesion and $\tau_{moving}$ between lesions then the rate of production of RecA filaments is given by the following formula:

Filament production rate = $\dfrac{1}{\tau_{stalled} + \tau_{moving}}$.

Further, the filament disassembly rate is $\dfrac{N_{fil}}{\tau_{RecA*}}$, where $N_{fil}$ is the number of RecA* filaments associated with the replication fork under consideration and $\tau_{RecA*}$ is the average persistence time of a RecA* filament.

Because the rates of filament production and disassembly are much faster than all other processes we are interested in (the transcription of SOS genes and the rate of NER repair) [13] we can assume that the number of RecA* filaments at any given time is such that the production rate equals the disassembly rate, i.e.,

$N_{fil} = \dfrac{\tau_{RecA*}}{\tau_{stalled} + \tau_{moving}}$.

The total amount of RecA*, $r^*$, is given by the above expression multiplied by $L_{RecA*}$ – the average length of a RecA* filament (taking into account the finite probability of forming a filament at each lesion a fork encounters) – and $N_f$, the total number of replication forks currently duplicating DNA in a cell, i.e.,

$$r^* = N_f L_{RecA*} \dfrac{\tau_{RecA*}}{\tau_{stalled} + \tau_{moving}}. \qquad (1)$$

After fixing the parameter values based on experimental data (see Methods) this relation gives a RecA* level of approximately 100 nM for a fixed lesion density produced by a UV dose of 2 $J/m^2$, while it gives more than 400 nM for a UV dose of 50 $J/m^2$ (this neglects the effects of Pol V which will be discussed later). The process shown in Fig. 2 is thus a simple way for the cell to "count" the number of lesions on the DNA using a "memory" which is the finite existence time of a RecA filament.

## *Accumulation and heterodimerization of UmuD'*

This is a short time memory lasting only for a time $\tau_{RecA*}$. However, the rate of UmuD' production is proportional to the amount of RecA*, therefore the UmuD' level is a measure of RecA* level integrated over time. Thus, UmuD' accumulates if damage (and therefore RecA*) persists for a long time. In our model, with RecA* at its maximum possible level, the timescale for the UmuD' level to exceed that of UmuD is around 15 minutes. For smaller UV doses, and therefore lower RecA*, this rise time can be more than 35 minutes. UmuD' is an integral component of the error-prone polymerase Pol V. However, UmuD' has to accumulate to a fairly high level before Pol V appears in any detectable quantities. The main reason for this is a strong physical interaction between UmuD and UmuD'. The binding between them is stronger than that between UmuD or UmuD' pairs; when UmuD and UmuD' are mixed in equimolar concentrations the heterodimer is found to be much more abundant than either homodimer (UmuD-UmuD and UmuD'-UmuD') [14]. This strong binding ensures that unbound UmuD' homodimers required for Pol V formation appear in sufficient quantities only when (and if) the total concentration of UmuD' exceeds that of UmuD.

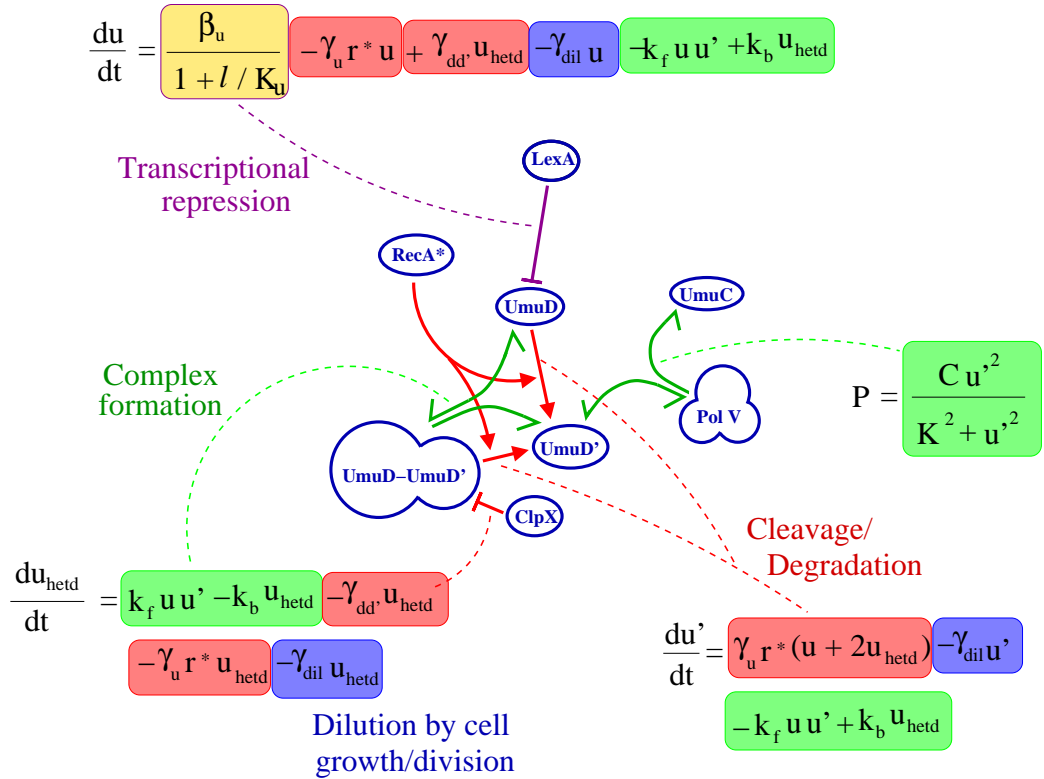

**Figure 3:** Mutagenesis subnetwork and associated dynamical equations (only those links are shown which have corresponding terms in the equations). $u$, $u'$ and $u_{hetd}$ are the concentrations of UmuD, UmuD' and UmuD-UmuD' heterodimer, respectively. $r^*$ is the RecA* level. $l$ is the LexA level. $k_f$, $k_b$, $\beta_u, K_u$, $\gamma_u$, $\gamma_{dd'}$, $\gamma_{dil}$, $C$ and $K$ are parameters (see Methods for their values). The equations describe 1) LexA repressed production of UmuD (highlighted in yellow), 2) RecA* mediated cleavage of UmuD (highlighted in red), 3) heterodimerization of UmuD and UmuD' (highlighted in green), 4) degradation of UmuD' by ClpX when in the heterodimer, releasing UmuD (red), 5) dilution of all proteins due to cell growth/division (indicated by blue) and 5) formation of Pol V (green).

Figure 3 shows the equations we use to model the dynamics of UmuD, UmuD' and Pol V. These equations model the following processes:

1. LexA represses the production of UmuD ($\beta_u, K_u$); here, we assume a Hill coefficient of 1 based on the fact that the upstream region of the UmuD promoter has only one LexA binding site [15].
2. RecA$^*$ catalyzes the intermolecular cleavage of UmuD [16] (of both free and heterodimer forms) to produce UmuD' at rate $\gamma_u$
3. UmuD and UmuD' form a heterodimer [14] with on- and off-constants given by $k_f$, and $k_b$.
4. ClpX degrades UmuD' (but not UmuD) when it is in the heterodimer [17], at rate $\gamma_{dd'}$.
5. All molecules are diluted by cell growth and division ($\gamma_{dil}$).

Pol V is composed of two units of UmuD' bound with one unit of UmuC protein. Thus, the level of Pol V cannot exceed that of UmuC ($C$), but for small amounts of UmuD' it is proportional to $u'^2$. $K$ controls how much of the UmuD' homodimer is required to saturate the levels of Pol V. The UmuC concentration $C$ for simplicity is assumed to be constant during the narrow time window where it matters (i.e., when $u'$ is non-zero).

The qualitative aspects of the dynamics produced can be understood by looking at a simplified version of these equations: Since RecA* levels change relatively slowly, first consider UmuD and UmuD' levels at a fixed RecA* concentration, and thus, a constant UmuD → UmuD' cleavage rate $\gamma_u r^*$. If the heterodimerization is extremely strong, the time course of the total (free + heterodimer) UmuD' ($u'_{tot} = u' + u_{hetd}$) satisfies the following rate equation (see Methods for the derivation from the equations in Fig. 3):

$$\frac{du'_{tot}}{dt} = \gamma_u r^* u_{tot} - \gamma_{dd'} \min(u_{tot}, u'_{tot}) - \gamma_{dil} u'_{tot} \qquad (2)$$

Here $u_{tot} = u + u_{hetd}$ is the total concentration of noncleaved UmuD (in free or heterodimer form). The first term, $\gamma_u r^* u_{tot}$, is the production of UmuD' due to the cleavage of UmuD, the second term is the ClpX-dependent degradation of UmuD' inside UmuD'-UmuD heterodimers, while the last term is the decrease in the concentration of UmuD' due to cell growth and division (the dilution term) common for all proteins in the cell. With LexA and RecA* levels fixed, i.e., $\gamma_u r^*$ constant, we can calculate the steady-state levels of UmuD and UmuD' from these equations, and hence, the condition for Pol V to be present, i.e., when UmuD' exceeds UmuD: $u'_{tot} > u_{tot}$. Setting $du'_{tot}/dt = 0$ and $\min(u_{tot}, u'_{tot}) = u_{tot}$ we obtain the condition for $u'_{tot} > u_{tot}$ in the steady state:

$$\gamma_u r^* > \gamma_{dd'} + \gamma_{dil} \qquad (3)$$

independent of UmuD production and degradation rates. Thus, Pol V abruptly appears once the RecA* level, and hence the value of $\gamma_u r^*$, crosses and stays above the required threshold for long enough to allow UmuD' to accumulate and pass the UmuD level. This analysis also suggests that there would be a threshold minimum UV dose below which Pol V does not appear because the NER repair brings down DNA damage quickly enough to bring the level of RecA* below the amount required to satisfy Eq. 3.

## *Feedback from Pol V to RecA**

The behavior of replication forks at lesions (described above) naturally provides a positive feedback from Pol V to RecA* because Pol V reduces the stall time at the lesion, $\tau_{stalled}$ (ref. [2] estimates that Pol V bypasses lesions with 100 to 150-fold higher efficiency than Pol III). This is illustrated in Figure 4: Initially, there is no Pol V. However, other "non-mutagenic" Trans Lesion Synthesis (TLS) polymerases, Pol IV and Pol II (DinB or PolB), which are always present in the cell, ensure that even in the absence of Pol V the stalled replication fork could still bypass a lesion [18] at a rate we denote $1/\tau_{stalled}^{(0)}$. In Fig. 4 this rate is slow enough that by the time the fork reaches the next lesion (after a time $\tau_{stalled}^{(0)} + \tau_{moving} > \tau_{RecA*}$) the first filament disassembles. At a later time, when Pol V appears, the stall time reduces dramatically [2, 19]. The scenario depicted in Fig. 4 assumes the bypass rate is dominated by Pol V-assisted bypass (for the more general treatment used in our model, see the Methods section). In this case, the reduction in stall time from $\tau_{stalled}^{(0)}$ to $\tau_{stalled}^{(PolV)}$ when Pol V appears is sufficient to allow the replication fork to reach a second lesion before the first RecA* filament disassembles. Therefore, the RecA* level rises when Pol V appears. When this rise is fast enough, which occurs for a large enough UV dose, this results in a second peak in LexA-controlled promoter activities, as shown in Fig. 5. Thus, the second peak is a natural consequence of the mechanism for setting RecA* levels represented by Eq. 1. This prediction of the model is confirmed by the recent single-cell fluorescence experiments of Friedman *et al.* [10]. They also found that the second peak was washed out when the signal was averaged over many cells. This is probably because of cell-to-cell variations. Among the parameters, which can vary between cells, is the number of replication forks. We find that averaging the LexA-controlled promoter activity predicted by our model over many cells with differing numbers of replication forks produces a curve with a single peak (Fig. 5, red dashed line) as observed in the experiments.

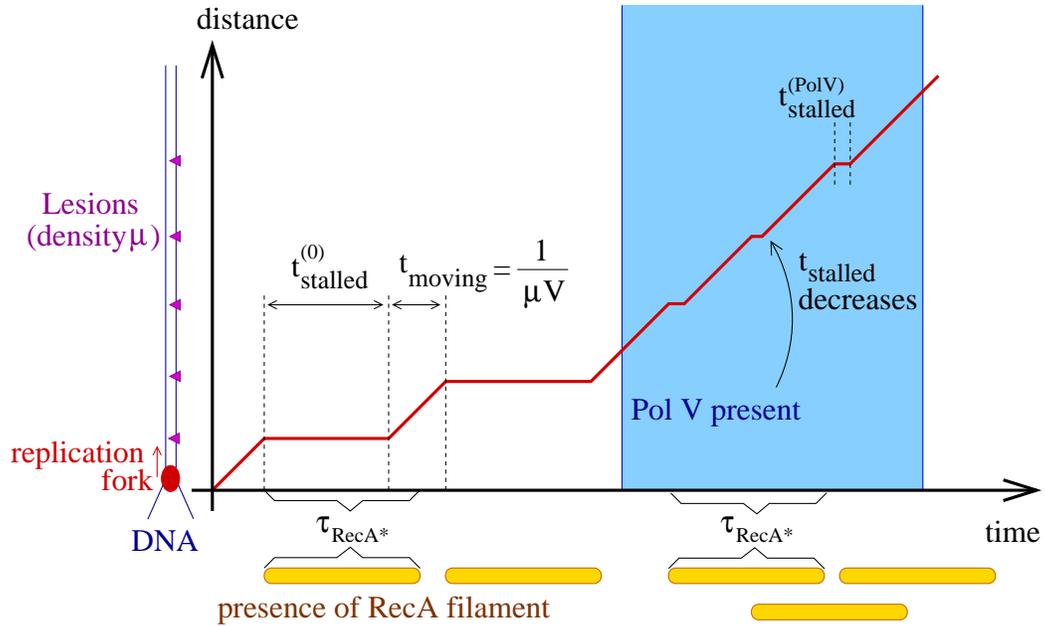

**Figure 4.** Mechanism of Pol V to RecA* feedback. The red line shows the progress in time of a replication fork on DNA with UV-induced lesions. Each yellow bar shows the time span of existence of the RecA filament formed when the fork encounters a lesion. In the absence of Pol V each RecA filament disassembles before the next one is created. At a later time, when Pol V appears (light blue region), the stall time is substantially reduced, so that a new RecA filament is created before the previous one disassembles, hence the level of RecA* goes up.

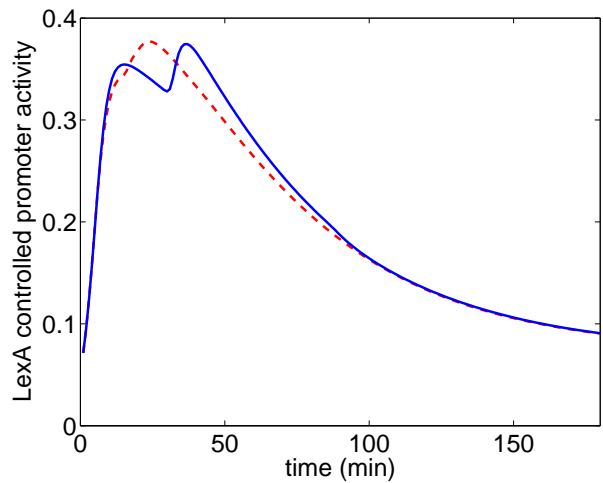

**Figure 5.** Activity of a hypothetical LexA-regulated promoter, parameterized as $\propto 1/(1+(LexA/100nM))$. The blue curve is for default parameters with UV dose $20 J/m^2$. The position of the second peak coincides with the generation of Pol V and is due to the feedback between Pol V and RecA* levels, mediated by Pol V- assisted bypass of replication forks. The red curve is produced by averaging over 200 hundred runs of the model, each with a different value for the number of replication forks, $N_f$, uniformly distributed between 1 and 3 ($N_f = 2$ for the blue curve).

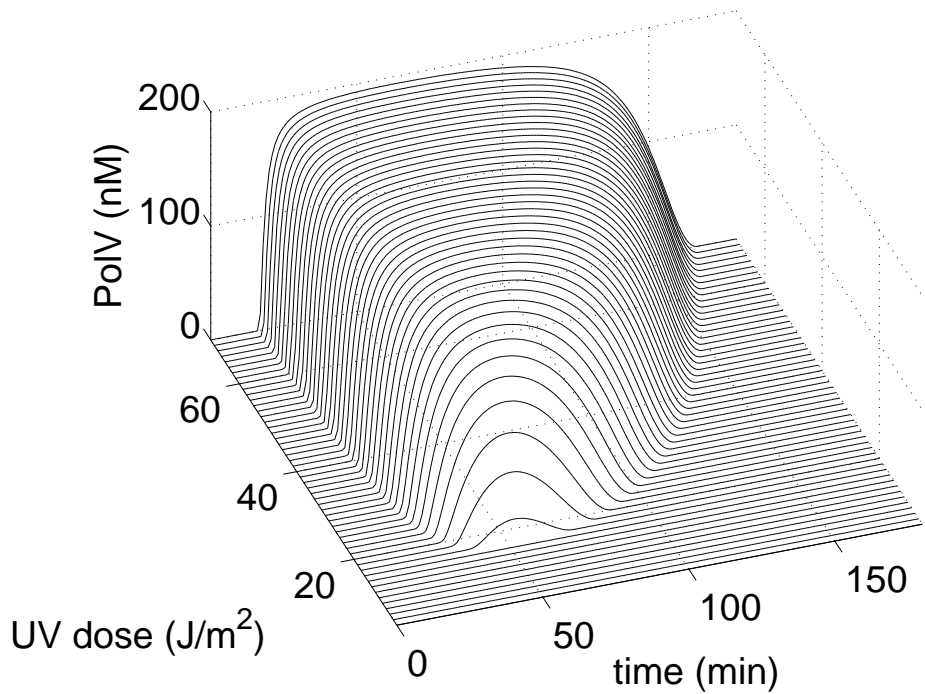

**Figure 6. Pol V concentration as a function of time, following an instantaneous pulse of UV at time zero, for different UV doses. Below a threshold of around 17 $J/m^2$ there is no Pol V, and hence, no mutagenesis. The sharp onset of Pol V, in relation to both time and UV dose, is a direct consequence of the strong heterodimerization of UmuD-UmuD'. Thus, mutagenesis occurs only if the amount of UmuD' exceeds that of UmuD, so that some free UmuD' is left to generate Pol V.**

## Tight control of Pol V levels

The model reveals an almost digital response of Pol V levels to UV, which provides very tight control of mutagenesis. Figure 6 shows the predictions of our model for the time course of Pol V (UmuD$'_2$C) for different UV doses. In these simulations, the cell is subjected to an instantaneous pulse of UV at the specified dose at time zero. The main features of this plot are

1) the existence of a UV dose (around 17 $J/m^2$) below which the Pol V level is very low. Thus, with low damage, mutagenesis is virtually absent and DNA repair is error-free.

2) a sharp onset in the generation of Pol V at around 15-35 mins for UV doses larger than 17 $J/m^2$. The time of onset is largely UV-independent at high doses.

3) a rapid turnoff of Pol V at variable times that increase with the UV dose.

This plot confirms several points suggested by the analysis of the model in the previous sections. Firstly, the existence of a minimum threshold UV dose below which no Pol V is produced is a consequence of the equations described in Figure 3 and, in particular, Eq. 3. The rapid onset and the later rapid decrease of Pol V is due to the combination of heterodimerization and the previously described positive feedback from Pol V to RecA* levels. We provide more evidence to support this conclusion in the next section.

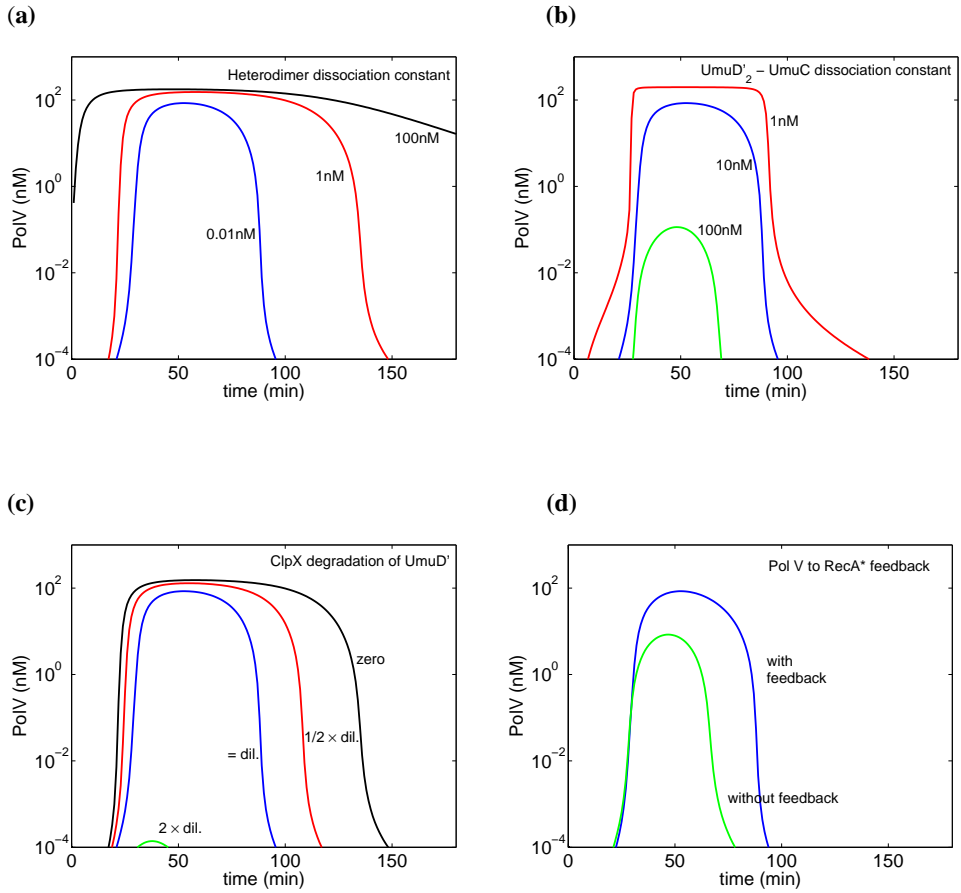

**Figure 7. Temporal profile of Pol V following a UV dose of $20 J/m^2$ as a function of various parameters. a)** The three curves refer, respectively, to strong ($K_{dd'} = 0.01$ nM, blue), medium ($K_{dd'} = 1$ nM, red), and weak ($K_{dd'} = 100$ nM, black) binding between UmuD and UmuD'. **b)** The effects of changing the binding constant between the UmuD' homodimer and UmuC: As binding strength $1/K$ in Eq. 12 increases ($K = 100, 10, 1$ nM for the green, blue, and red curves, respectively), the Pol V concentration saturates at the 200nM value set by the maximum cellular level of UmuC. **c)** For strong binding ($K_{dd'} = 0.01$ nM), the three curves show the effect of increasing the degradation rate $\gamma_{dd'}$ of UmuD' by ClpX. As a default, the degradation rate is set equal to the dilution rate $\gamma_{dil}$ (blue). The rate is half of the dilution rate for the red curve, whereas it is zero for the black curve. For the green curve, the degradation rate is double that of the dilution rate, which, at this level of UV damage, results in almost no Pol V. **d)** The effect of removing the Pol V to RecA* feedback. The blue curve is when there is feedback (as in the other figures). The green curve is when there is no feedback, i.e., $\tau_{stalled} = \tau_{stalled}^{(0)}$ irrespective of the Pol V level.

## The role of association strengths, degradation and feedback on onset and turnoff of mutagenesis

The above analysis uses a simplifying assumption that the binding between UmuD and UmuD' is infinitely strong so that the level of UmuD-UmuD' heterodimer is simply given by min([UmuD'],[UmuD]). The model can be used to examine the importance of the strength of this interaction in the mutagenesis response. Figure 7a illustrates the effect of decreasing this dissociation constant ($K_{dd'}$). It shows that a strong association is critical in setting the abruptness and positions of both the turn-on and turn-off points for Pol V. Another relevant protein-protein interaction is the binding between UmuC and UmuD' homodimers to form Pol V ($K$). This is one of the parameters for which experimental data is not available (see Methods). However, figure 7b shows that decreasing this dissociation constant makes the Pol V profile more "digital", i.e., more step-like with the concentration being either zero or maximum most of the time. Decreasing the ClpX-dependent degradation rate of UmuD' in the heterodimer, $\gamma_{dd'}$, mostly delays the turnoff of Pol V without affecting its turn-on time (Figure 7c).

Figure 7d shows the effect of turning off the positive feedback from Pol V to RecA*. Clearly, this feedback, combined with strong heterodimerization, is a crucial ingredient in the rapid onset of Pol V. Without feedback, the Pol V level is an order of magnitude lower compared to when there is feedback.

## Saturation of RecA* and peak activity levels

Another direct implication of Eq. 1 is that the peak amount of RecA* saturates as the UV dose is increased. Indeed, as the density of lesions $\mu$ rises, $\tau_{moving} = 1/\mu v$ decreases. According to Eq. 1, the RecA* level saturates once $\tau_{moving}$ becomes much smaller than $\tau_{stalled}$. Consequently, the height of the first peak of LexA-controlled promoter activity eventually saturates at UV high doses. During the second peak of promoter activity, the RecA* concentration rises again as $\tau_{stalled}$ drops due to the Pol V - assisted bypass of lesions. The height of the second peak also saturates, but at higher UV doses. For the parameters used in our model, the amplitude of the first peak of promoter activity reaches 90% saturation around 25 $J/m^2$, while that of the second peak – around 48 $J/m^2$ (see Figure 8b.) This prediction of our model is in agreement with the experimental data in Fig. 4C of Ref. [10], which shows that the saturation of the second peak occurs at a higher UV dose than for the first peak. However, that data shows the peak height averaged over a cell population. Therefore, in order to compare our model directly to the data, we show in Fig. 8a, the peak heights averaged over 200 runs with varying $N_f$. The resultant peak height versus UV dose curves match the data of Ref. [10] satisfactorily with the exception of the first peak data point at 50 $J/m^2$ which is *lower* than the previous data points. One explanation could be the ambiguity in the averaging procedure because, especially at higher UV doses, the second peak may sometimes be large enough to swamp out the first one, and hence be counted as a first peak, thus raising the red curve. Note, however, that at the single-cell level our model will always show a monotonically increasing peak height as UV dose is increased.

The behavior of our model also agrees with Fig. 4A of ref. [10] from which we conclude that the second peak of promoter activity starts to appear at a considerable frequency for UV-doses between 10 and 20 $J/m^2$. The threshold of 17 $J/m^2$ predicted by our model (the same as the threshold for mutagenesis) is consistent with this.

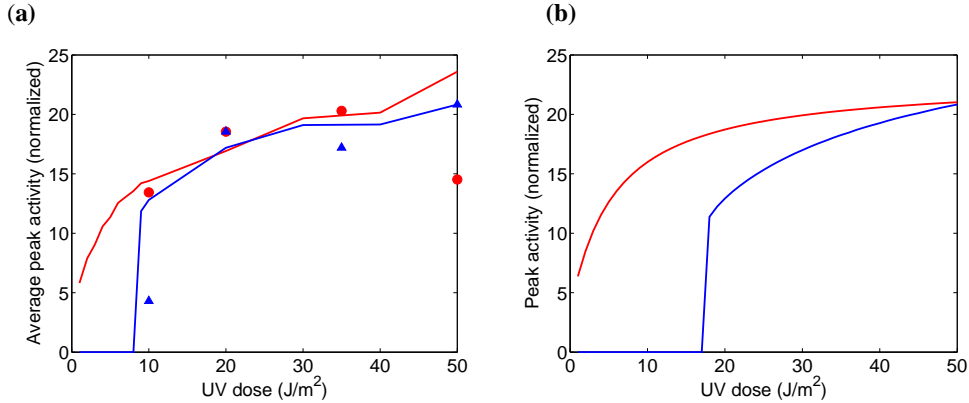

**Figure 8.** (a) Height of the first (blue curve) and second (red curve) peaks in LexA-regulated promoter activity, averaged over 200 runs each with a different value of $N_f$, uniformly distributed between 1 and 3 (other parameters remained fixed as in Fig. 4) as a function of the UV dose. Red circles and blue triangles show the corresponding data from Fig. 4C of Ref. [10]. To facilitate comparison, the height of the red (blue) curve was normalized to match the maximum experimentally observed peak activity. (b) The peak heights as a function of UV dose for a single run with $N_f = 2$, with the same normalization as in (a). The saturation of peak heights for increasing UV doses in our model is a consequence of Eq. 3.

## *The response of the cell to continuous UV damage*

The SOS response of bacteria to radiation is typically studied by exposing them to a very short burst of UV light and then following the repair of the DNA damage. However, in environments for which bacteria are evolutionarily adapted, there may be both short bursts of the UV radiations, similar to the experimental conditions imposed on them, as well as much longer spells of low intensity UV exposure. The latter type of perturbation might not be well suited for *in-vivo* experiments but is easily achievable in our *in-silico* model. Adding a new term representing a continuous low-rate of production of lesions (see Methods), gives rise to a stable steady state wherein the rate of NER repair equals the rate of creation of the new DNA damage. Figure 9a shows the typical response to a continuous UV dose which is low enough that mutagenesis is never triggered; LexA and RecA* take around 60 minutes to reach a steady state. Experiments in which cells were exposed to continuous UV damage because of the presence of a constant amount of mitomycin C also indicate that the SOS response (rates of LexA repressor synthesis and cleavage) took 60 minutes to reach a steady state [3], confirming this prediction of our model.

We also simulated the response of our virtual cell to a pulse of UV-radiation of a given integral intensity and duration that varied from 0 to 300 minutes. Figure 9b separates the mutagenic and non-mutagenic regions of parameter space. Initially, the magnitude of the SOS response weakly increases with prolonging the duration of the pulse. This response was expected since very short pulses give the NER subsystem time to repair some lesions before replication forks encounter them; therefore, the average RecA$^*$ concentration is less than for slightly longer pulses. The threshold for activating the mutagenesis subsystem reaches its minimum value for ~ 60 minute pulse, and then it increases linearly throughout the duration of the pulse indicating that the cell has reached the steady state in which mutagenesis is not triggered by the total intensity of the pulse, but rather by a sufficiently high rate of production of new lesions corresponding to a UV intensity per unit time of about 1.5 $mW/m^2$. This is an order of magnitude less than the typical solar UV intensity of 7-10 $mW/m^2$ in Copenhagen at noon on a clear day in December; for comparison, the solar UV intensity in the tropics in similar conditions is over 100 $mW/m^2$ (see http://www.temis.nl/uvradiation/UVindex.html).

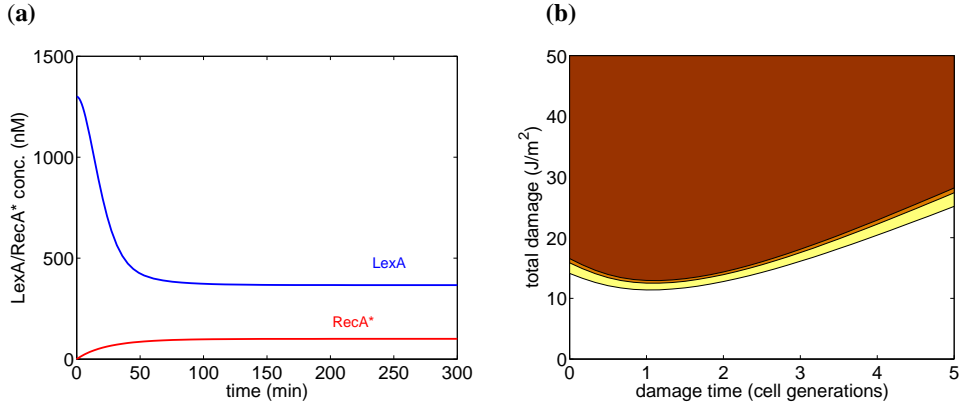

**Figure 9.** SOS response to prolonged UV exposure. a) LexA (blue) and RecA* (red) levels as a function of time in the presence of a continuous source of UV, which in 300 minutes (i.e., 5 cell generations) produces as many lesions as an instantaneous pulse of 20 $J/m^2$. b) The presence or absence of mutagenesis in our model in response to a pulse of UV radiation of a given integral intensity (y-axis) and duration (x-axis). Mutagenesis was detected in the colored regions. The criterion for its detection was the Pol V level crossing a specified threshold: 0.1 nM (yellow region), 1 nM (orange region), 10 nM (brown region).

## Discussion

When bacteria experience a large amount of DNA damage their response has a mutagenic component which, it has been suggested, might afford some evolutionary advantage by altering the genome of offspring that would allow some of them to better survive high levels of the damage-inducing agents [20]. Precursors to an error-prone polymerase have also been implicated in slowing down DNA replication [21], thereby allowing additional time for accurate repair processes to remove lesions from the DNA. This delay is immediately terminated once the error-prone polymerases are fully formed. However, this kind of evolutionary strategy would be harmful where there was no damage, or when it was sufficiently low that it could be quickly repaired by error-free mechanisms. Hence, mutagenesis must be tightly regulated.

The main features of the mutagenic component of the SOS response system, according to published literature, are the following:

1) Mutagenesis is characterized by a sharp temporal onset and turnoff and threshold-like behavior as a function of UV dose. There is a strong experimental evidence for this: For example, Rangarajan *et al.* [18], observed that in the absence of Pol II masking the effects, Pol V - assisted bypass rapidly appears around 45 minutes after the irradiation. Also, from Fig. 4 of Ref. [21] we may conclude that the UmuD' concentration becomes comparable to that of UmuD around 30 mins after irradiation, irrespective of UV dose. This exactly matches the time at which Pol V appears in our model when UmuD-UmuD' binding is very strong.

2) Mutagenesis gives rise to the second peak in activity of the SOS regulon. This is inferred from data in Ref. [10] that shows this second peak is absent in mutants which lack UmuD or contain an uncleavable version of it.

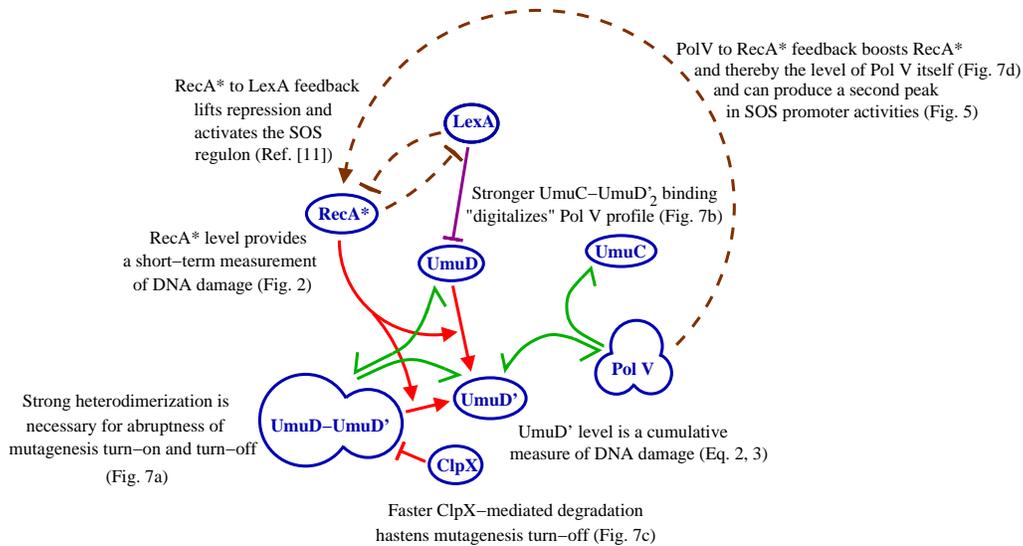

**Figure 10. Summary of the main lessons of our model of the SOS response and mutagenesis.**

We constructed a network model of mutagenesis in the bacterial SOS response system to account for these features. Figure 10 summarizes the key aspects of the behaviour of the system that emerged in our simulations. We demonstrated that strong binding between UmuD and UmuD' is necessary for the sharp onset of mutagenesis and for its turn off when UmuD' again falls below UmuD (see Fig. 7a). Thus, initially, when levels of UmuD' are low, almost all of the UmuD' is sequestered in heterodimers so that no Pol V is generated. However, UmuD' is being constantly produced by the cleavage of UmuD, whose production, in turn, is elevated due to the de-repression of its promoter. If the UV damage is large enough, eventually the concentration of UmuD' rises sufficiently to exceed that of UmuD and allow the formation of Pol V. Additional control is afforded by the degradation of the UmuD-UmuD' heterodimer by ClpX, which removes UmuD' while freeing UmuD for further cleavage or dimerization. Although this degradation is not essential for the system's qualitative behavior it substantially influences the turnoff time and rate (Fig. 7b). Indeed, without it, turnoff could be only realized by the reduction in UmuD cleavage rates due to DNA repair, and would depend solely on the slower NER mechanism. In addition, Lon actively degrades UmuD homodimers and UmuC [17]; its physiological advantages are unclear. Including this mechanism in our model does not affect the system's qualitative behavior, provided the degradation rate is not too large.

We suggested a simple mechanism by which the RecA* level can serve as a measure of the lesion density (see Eq. 1). This mechanism relies on the possibility for RecA filaments to exist for some finite time after the replication fork has bypassed the lesion where the filament was created (note that we assume that this happens whether the lesion was on the leading or lagging strand). This allows the replication fork to sample a stretch of DNA, thus making a measurement of the damage density which is then manifested in the RecA* level. A direct implication of this mechanism is that there is a positive feedback from the Pol V to RecA* levels (see Fig. 4). The resulting temporary increase in RecA* levels due to the sudden appearance of Pol V is exactly the ingredient required to explain the resurgence of the SOS response 30-40 minutes after irradiation, observed in the single-cell experiments of Friedman *et al.* [10]. In addition, this mechanism also explains their observation of saturation of the peak promoter activities, and hence RecA* levels, upon increasing the UV dose (see Fig. 8). Note that the first peak in promoter activity is produced due to changes in the lesion density, and thereby $\tau_{moving}$, as NER swings into action, while the second peak is due to changes in $\tau_{stalled}$, due to the action of Pol V. $\tau_{moving}$ and $\tau_{stalled}$ both affect RecA* level in the same way, being symmetrically placed in the denominator of Eq. 1, but are influenced by different mechanisms.

Of course, various parameters that we use in our model will vary from cell to cell in a population. Such stochasticity plays an important role in the observed behavior probably only for those components that are present in low numbers in the cell. Therefore, we consider that stochasticity in the number of replication forks is likely to be the most important source of cell-to-cell variability for the SOS system. As a default we take this number, $N_f$, to be 2. However, for comparing with data obtained from cell populations, we averaged over several runs where $N_f$ was allowed to vary between 1 and 3 (see Fig. 4 and 8). Another component present in a relatively low concentration is UmuC, variation of which is shown in Fig. 7b. Fig. 7c shows that the Pol V profile is quite sensitive to ClpX. Therefore, this might be another source of variability.

As more directly observable predictions of our model, we offer the following:
(i) Overexpression of ClpX should considerably reduce the Pol V concentration. At the other extreme, the absence of ClpX would lead to Pol V being turned-off at a later time than in wild type cells (see Fig. 7c).
(ii) Overexpression of UmuC results in a flatter Pol V profile (see Fig. 7b), while an UmuC mutant should not be able to produce Pol V and hence should behave like the $\Delta$UmuDC and un-cleavable UmuD mutants studied in Ref. [10].
(iii) We find that some overexpression of UmuD (upto a factor 2), or the introduction of more UmuD before the UV pulse, causes an increase in the second peak height. Further, the peak occurs earlier, sometimes even swamping the first peak. However, removing the LexA repression of UmuD (say by introducing UmuD on a plasmid with an unregulated promoter) results in the vanishing of the second peak, except at particularly high UV doses, because UmuD' is not formed fast enough to cross the UmuD level.

There are alternative mechanisms by which UmuD and UmuD' could affect LexA levels within the framework of the SOS model considered here. We discuss two below which, unlike the mechanism we have concentrated on so far, could produce a second peak by causing a temporary rapid *decrease* of LexA, i.e. a *trough* in the average promoter activity profile.

(a) UmuD competes with LexA for the RecA$^*$ binding sites. Conceivably, high UmuD levels could prevent the access of LexA to them, thereby reducing its cleavage rate. This would create a trough (not a peak!) in LexA-controlled promoter activity at the peak of UmuD concentration. An important observation in Ref. [10] requires this mechanism of competition: peak LexA-controlled promoter activity in the $\Delta$UmuDC mutant appears to increase with increasing UV dose (in the range 20-35 $J/m^2$) rather than saturating, as in wild-type cells. Because of the absence of competition in the $\Delta$UmuDC mutant LexA is cleaved more and falls to a lower concentration than in wild-type cells, which, in turn, leads to a higher peak activity level of LexA-repressed promoters. However, this mechanism cannot alone be the whole story because it cannot explain the observation of Ref. [10] that cells lacking UmuD protein and those with only un-cleavable UmuD have the same phenotype that *does not* exhibit the second peak. Indeed, the latter would be better equipped to compete for LexA binding sites than the wild-type cells (since the uncleavable UmuD K97A retains the ability to bind to the RecA filament [22]). Thus, if this was the right explanation, they should exhibit an even stronger second peak of promoter activity. The fact that $\Delta$UmuDC and un-cleavable UmuD have the same phenotype strongly implicates UmuD' as the main factor responsible for generating the second peak.

(b) UmuD was shown to preferentially bind to the beta-clamp subunit of DNA polymerase III while UmuD' prefers to bind to the epsilon subunit [12]. Thus, by sequestering the beta-clamp, large levels of UmuD could possibly reduce the processivity of Pol III, which feeds back onto RecA* and LexA levels via $\tau_{moving}$ in Eq. 1. Once again, the effect of this would be to produce a trough in LexA controlled promoter activity, since high levels of the UmuD would lower the processivity of Pol III and hence prolong $\tau_{moving}$ that, in turn, would lead to a drop in RecA$^*$ levels. Then, later when UmuD levels drop and the processivity of Pol II increases it could lead to a second peak.

To explore these postulated causes of the second peak in RecA$^*$ levels, we incorporated each of these two feedback mechanisms into our model. In the absence of Pol V $\to$ $\tau_{stalled}$ feedback, these mechanisms, alone or in combination, we found it difficult to generate the second peak in promoter activity at a reasonable time (within 40-50 min) after irradiation. However, their presence did not interfere with the manifestation of this feature when the Pol V $\rightarrow \tau_{stalled}$ feedback was included in the model. Therefore, they might well be operating in parallel in cells.

Friedman *et al.* also reported the existence of a third peak in the LexA-controlled promoter activities [10]. Unlike the second peak, the third one exhibited less fluctuations between individual cells. Indeed, its existence was previously mentioned in Ronen *et al.*'s study which used a signal averaged over many cells [23]. The generation of this third peak requires a mechanism that would increase the amount of RecA$^*$ at around 100-120 minutes after UV irradiation. As Friedman *et al.* suggest, one possible candidate is DinI, an SOS gene that is also repressed by LexA and induced in response to DNA damage. DinI is known to (a) stabilize already formed RecA$^*$ filaments, (b) prevent RecA$^*$-mediated cleavage of UmuD, and, (c) leave the RecA$^*$-mediated cleavage of LexA unaffected [22, 24]. The first property would cause an increase in $\tau_{RecA*}$ and thus, if DinI were to be generated, or become sufficiently active, around 100-120 min after the initial damage, it would result in a rise in RecA$^*$ levels and a new peak in LexA-dependent promoter activity. Ref. [25] suggests that DinI coats RecA$^*$ filaments with a 1:1 stoichiometry. Therefore, its activity would become substantial only when its levels exceeded the RecA$^*$ levels in the cell. Since the RecA levels are high ($\sim$ 7200 [4]) even in the absence of damage, and increase further due to de-repression of the SOS regulon, it might take up to 100-120 minutes until the induced DinI levels would overtake the diminishing levels of activated RecA$^*$. This speculation is corroborated by co-IP results presented in Fig 2B of Ref. [26]. A plausible evolutionary role for such delayed RecA$^*$ stabilization is that it would support a stable low-level SOS response when the cell is exposed to a persistent source of DNA damage.

Overall, our model systematizes causes and effects in the best known parts of the SOS response system in *E. coli*. It provides a framework for asking new questions about how (and why) the SOS response is organized. For instance, why is mutagenesis first initiated at such a late stage when only 10-20 percent of the original lesions remain untouched by NER? One hypothesis would be that mutagenesis is triggered in response to the presence of particular types of lesions which are less efficiently bypassed by other mechanisms. This can be tested by extending the model to incorporate different kinds of lesions with different feedback to the stall time. Another key aspect is the length of the time window during which Pol V is active, which is set by the degradation times of UmuD, UmuD' and UmuC. A more accurate determination of the regulation of these degradation times may shed light on effects of memory and why mutagenesis is at all initiated for severe DNA damage. In discussing this one should however keep in mind that mutagenesis may be designed to work primarily under a continuous source of DNA damage, and that the timing effects that we use to gain insight into the dynamics of the SOS response may be of secondary importance under typical real-world stresses. In any case, our study of the mutagenesis sub-network suggests that its behavior is quite "digital", in the sense that it makes a very quick transition from a state where there is no mutagenesis to a state where Pol V is fully activated.

# Methods

## The model

Our model of the SOS response defined by equations 4-12, below, builds upon excellent earlier work by Aksenov [11]. Equations 4, 6 and 7 are very similar to their counterparts in Ref. [11], while equations 5 and 8-12 are the new ones that we propose to describe mutagenesis and its feedback onto RecA* and LexA levels.

### RecA-LexA feedback

The dynamics of LexA ($l$) level is modeled using the following equation:

$$\frac{dl}{dt} = \frac{\beta_l}{1 + l/K_l} - \gamma_l r^* l - \gamma_{dil} l. \qquad (4)$$

Here, the first term models the self-repression of LexA production. We assume a Hill coefficient of 1 [23]. The second term is for the cleavage of LexA by RecA$^*$ (whose level is denoted by $r^*$), while the third term is the degradation of LexA in non-irradiated cells.

The RecA$^*$ level ($r^*$), in turn, is described by

$$r^* = N_f L_{RecA*} \frac{\tau_{RecA*}}{\tau_{stalled} + \tau_{moving}}, \qquad (5)$$

which is exactly the same as equation 1 in the main text. In writing this equation for the RecA* level we assume that the timescales involved in filament assembly and disassembly [13] are much smaller than those of transcriptional regulation and NER repair; therefore, a differential equation is not needed to describe the dynamics of RecA*.

Denoting the density of lesions by $\mu$ and the speed with which Pol III moves on undamaged DNA by $v$, we get the following expression for $\tau_{moving}$:

$$\tau_{moving} = \frac{1}{\mu v}. \qquad (6)$$

The density of lesions is not a constant; they are continuously being repaired by the NER mechanism, which we model as follows:

$$\frac{d\mu}{dt} = -\lambda \mu. \qquad (7)$$

Here we assume that the repair is limited by the number of lesions and not by the Uvr proteins, hence the repair rate is proportional to the lesion density. In taking the rate of repair per lesion, $\lambda$, to be a constant we ignore the feedback from LexA to mRNA production from *uvr* genes; following [11] we assume that the repair is limited by UvrC, which is not repressed by LexA.

## Mutagenesis and the feedback from RecA$^*$ and LexA

We can write the overall lesion bypass rate $1/\tau_{stalled}$ as the sum of two rates, bypass due to Pol V ($1/\tau_{stalled}^{(PolV)}$), and bypass due to all other mechanisms ($1/\tau_{stalled}^{(0)}$).

$$\frac{1}{\tau_{stalled}} = \frac{1}{\tau_{stalled}^{(0)}} + \frac{1}{\tau_{stalled}^{(PolV)}} \frac{P}{K_P + P}. \qquad (8)$$

Here $P$ is the Pol V level and $P/(K_P + P)$ is the factor describing the concentration dependence of the Pol V-assisted bypass rate. This generalizes the situation described in Fig. 4.

To describe the dynamics of the Pol V concentration we must first consider concentrations of free UmuD ($u$), free UmuD' ($u'$) and the UmuD-UmuD' heterodimer $u_{hetd}$:

$$\frac{du}{dt} = \frac{\beta_u}{1 + l/K_u} - \gamma_u r^* u + \gamma_{dd'} u_{hetd} - k_f u u' + k_b u_{hetd} - \gamma_{dil} u \qquad (9)$$

$$\frac{du'}{dt} = \gamma_u r^* [u + 2 u_{hetd}] - k_f u u' + k_b u_{hetd} - \gamma_{dil} u' \qquad (10)$$

$$\frac{du_{hetd}}{dt} = k_f u u' - k_b u_{hetd} - \gamma_{dd'} u_{hetd} - \gamma_u r^* u_{hetd} - \gamma_{dil} u_{hetd} \qquad (11)$$

Finally, the Pol V level, which feeds back to RecA$^*$ and LexA levels via the equation 8, is given by

$$P = \frac{C u'^2}{K^2 + u'^2}. \qquad (12)$$

These are the equations shown and explained in Fig. 3.

## Parameter values

Our model is fully specified by 18 parameters. Below we divide them into two groups, the first whose values we could fix directly or indirectly from published literature, the second for whose values there is inconclusive or no published data.

**Parameters fixed from experimental data:**

1. The repair rate by NER: $\lambda = 0.035 \text{ min}^{-1}$, corresponding to a half-life of approximately 20 min as reported in ref. [27] for cyclobutane pyrimidine dimers.

2. LexA concentration required for half-repression of LexA promoter: $K_l = 270 \text{ nM}$, corresponding to an induction ratio of approximately 5.8 (the relation between the two is $K = 1300/(I-1)$, where 1300 nM is the LexA level in undamaged cells [4]), interpolated from induction ratios of 6.7 and 4.8, measured at 30 and 42 degrees Celsius, respectively [15].

3. LexA concentration required for half-repression of UmuD promoter: $K_u = 60 \text{ nM}$, corresponding to an induction ratio of approximately 22.7, interpolated from induction ratios of 28 and 17 measured at 30 and 42 degrees Celsius, respectively [15].

4. Dilution rate: $\gamma_{dil} = \ln(2)/60 = 0.012 \text{ min}^{-1}$, estimated average from a scatter plot of cell doubling time in Ref. [10]. Further, 60 min is the reported half-life of LexA in non-irradiated cells treated with chloramphenicol, i.e., in the absence of production of LexA [4], as well as from pulse-labeling measurements of LexA cleavage rates [3].)

5. Speed of Pol III on undamaged DNA: $\nu = 1000 \text{ bp s}^{-1}$ [28].

6. UmuC level: $C = 200 \text{ nM}$ [29].

7. Number of lesions per unit UV dose: 50 per $J/m^2$. This corresponds to 250 lesions per *E. coli* genome for a dose of 5 $J/m^2$ [30]. Thus, for *E. coli* the initial lesion density is given by $\mu(t=0) = 10^{-5} \times D$, where $D$ is the UV dose in $J/m^2$.

8. Stall time in absence of Pol V: $\tau_{stalled}^{(0)} = 0.22 \text{ min}$. In Ref. [10], the height of the first peak in LexA-repressed promoter activities reaches half-maximum at around 7-8 $J/m^2$. In our model, this occurs when $\tau_{stalled} = \tau_{moving}$. At 7-8 $J/m^2$, $\mu = 7 - 8 \times 10^{-5}$ hence $\tau_{moving} \approx 0.22 \text{ min}$ and $\tau_{stalled} \approx \tau_{stalled}^{(0)}$ because the Pol V level is negligible. The chosen value is also consistent with an estimate of 10-12 seconds based on measurements of DNA synthesis in irradiated cells [30].

9. Stall time in presence of Pol V: $\tau_{stalled}^{(PolV)} = 0.022 \text{ min}$. Ref. [2] estimates that Pol V bypasses lesions with 100 to 150-fold higher efficiency than Pol III. However, *in vivo* the ratio of $\tau_{stalled}^{(0)}$ to $\tau_{stalled}^{(PolV)}$ will be smaller than that because, apart from Pol III, other polymerases like Pol II also contribute to the bypass rate [18]. Ref. [19] finds that in $\Delta umuDC$ cells, the frequency of replication past *cis-syn* T-T dimers, produced by 4 $J/m^2$ irradiation, is approximately 40 times lower than in cells containing a chromosomal copy of *umuDC*. In our simulations, we find that as long as $\tau_{stalled}^{(0)} \gg \tau_{stalled}^{(PolV)}$, the dynamics do not depend much on the value of $\tau_{stalled}^{(PolV)}$; therefore we conservatively chose $\tau_{stalled}^{(0)} / \tau_{stalled}^{(PolV)} = 10$.

10. Parameter determining RecA* levels: $N_f L_{RecA*} \tau_{RecA*} = 110 \text{ nM min}$.

11. LexA cleavage rate (per nM of RecA*): $\gamma_l = 8.8 \times 10^{-4} \text{ nM}^{-1} \text{min}^{-1}$. These two parameters are together fixed so that the maximum rate of LexA degradation is $\frac{N_f L_{RecA*} \tau_{RecA*}}{\tau_{stalled}^{(0)}} \times \gamma_l \approx 0.44 \text{ min}^{-1}$, corresponding to a half-life of around 1.5 min, chosen to match pulse-labeling measurements of LexA degradation rates in irradiated cells [3] and measurements in cells where LexA production was prevented by adding chloramphenicol [4]. Note that in our equations only the product of these two parameters, $N_f L_{RecA^*} \tau_{RecA^*} \times \gamma_l$, appears, so for the model simulations the individual value of each parameter is irrelevant as long as the product is preserved. However, we chose $N_f L_{RecA^*} \tau_{RecA^*} = 110$ nM min to fix the maximum possible RecA* level, $\frac{N_f L_{RecA^*} \tau_{RecA^*}}{\tau_{stalled}^{(0)}}$, to 500 nM. This is a reasonable number, obtained by assuming that $N_f = 2$ and the maximum RecA* level is achieved when both these replication forks are stalled. Then, with each fork leaving an ssDNA gap of 900 nucleotides [11] and given that each RecA* filament has 1 monomer per 3-5 nucleotides [4, 5] we obtain a maximum RecA* level around 500 nM.

12. UmuD cleavage rate (per nM of RecA*): $\gamma_u = 1.8 \times 10^{-4}$ nM$^{-1}$min$^{-1}$. This was chosen to be approximately 5 times slower than the LexA cleavage, estimated from the following data: After a UV dose of 20 $J/m^2$, the half-life of the UmuD to UmuD' cleavage is approximately 45 min [12], while at the same UV dose, ref. [4] reports a half-life for LexA around 7-10 min.

13. Maximal LexA production rate: $\beta_l = 86$ nM min$^{-1}$ (chosen so that in undamaged cells the level of LexA stabilizes to 1300 nM [4] for which it must satisfy the formula $\beta_l = 1300 \times \gamma_{dil} \times (1 + 1300/K_l)$.)

14. Maximum UmuD production rate: $\beta_u = 47$ nM min$^{-1}$ (chosen so that in undamaged cells the level of UmuD stabilizes to 180 nM [29] for which it must satisfy the formula $\beta_u = 180 \times \gamma_{dil} \times (1 + 1300/K_u)$.)

15. Constant involved in concentration dependence of Pol V dependent bypass: $K_P = 10$ nM. The level of Pol V in irradiated cells ranges from 15 to 60 molecules for small UV doses [31]. We chose $K_P$ such that this range of Pol V level will produce a substantial, but not saturating, contribution to $\tau_{stalled}$.

**For parameters where data was unavailable or inconclusive, we have scanned a range of values around the following chosen defaults:**

16. Binding constant of the UmuD-UmuD' heterodimer: $K_{dd'} = k_b / k_f = 0.01$ nM, chosen to be very strong ($k_b$ and $k_f$ were individually chosen to be relatively large so that the heterodimer was, in practice, always in equilibrium with unbound UmuD and UmuD'.)

17. Degradation of UmuD-UmuD' by ClpX: $\gamma_{dd'} = \gamma_{dil} = 0.012$ min$^{-1}$.

18. UmuD' level required for Pol V level to reach half maximum: $K = 10$ nM.

## Initial conditions

For initial conditions, we have used the experimentally reported levels in wild-type cells: LexA=1300 nM [4], RecA*=0 (naturally existing ssDNA, e.g., lagging strand replication gaps, does not activate RecA in the absence of DNA damage [4]), UmuD=180 nM and UmuD'=0 [29].

## The limit of strong heterodimerization

First, adding equation 9 and 11 we get

$$\frac{du_{tot}}{dt} = \frac{\beta_u}{1 + (l/K_u)^{h_u}} - \gamma_u r^* u_{tot} - \gamma_{dil} u_{tot}. \quad (13)$$

Here, $u_{tot} = u + u_{hetd}$ is the total amount of UmuD. Similarly, adding equations 10 and 11 and using the fact that when heterodimer binding is infinitely strong, $u_{hetd} = \min(u_{tot}, u'_{tot})$, we get equation 2 in the main text

$$\frac{du'_{tot}}{dt} = \gamma_u r^* u_{tot} - \gamma_{dd'}\min(u_{tot}, u'_{tot}) - \gamma_{dil} u'_{tot} \qquad (14)$$

*Model of prolonged exposures to low-level UV radiation*

To model the dynamics during exposures to pulses of UV radiation of finite (possibly long) duration we modified the equation 7 as follows:

$$\frac{d\mu}{dt} = s - \lambda\mu, \qquad (15)$$

where $s$ is a source term, which is a non-zero constant when $0 \le t \le t_d$. Here, $t_d$ is the duration of the UV pulse and $s \times t_d$ is the total integral UV dose.

# Acknowledgements

We thank Joel Stavans, Nir Friedman, Penny Beuning, Avril Woodhead, Richard Setlow, Ian Dodd and anonymous referees for very useful discussions and critical comments on the manuscript. Work at Brookhaven National Laboratory was carried out under Contract No. DE-AC02-98CH10886, Division of Material Science, U.S. Department of Energy. SK and KS thank the Theory Institute for Strongly Correlated and Complex Systems at BNL for financial support during visits to BNL when some of this work was accomplished. Work at the NBI was supported by the Danish National Research Foundation.

# References

1. Walker, G.C., *Chapter 89: The SOS Response of Escherichia coli*, in *Esherichia coli and Salmonella typhimurium, Vols I and II.*, J. Ingraham, et al., Editors. 1987, American Society of Microbiology: Washington, DC.
2. Tang, M., et al., *UmuD'2C is an error-prone DNA polymerase, Escherichia coli pol V.* Proc Natl Acad Sci U S A, 1999. **96**: p. 8919-8924.
3. Little, J.W., *The SOS Regulatory System: Control of its State by the Level of RecA Protease.* J. Mol. Biol., 1983. **167**: p. 791-808.
4. Sassanfar, M. and J.W. Roberts, *Nature of the SOS-inducing Signal in Escherichia coli - The Involvement of DNA Replication.* J Mol Biol, 1990. **212**: p. 79-96.
5. Takahashi, M. and B. Norden, *Structure of RecA-DNA Complex and Mechanism of DNA Strand Exchange Reaction in Homologous Recombination.* Adv. Biophys., 1994. **30**: p. 1-35.
6. Little, J.W., et al., *Cleavage of the Escherichia coli lexA protein by the recA protease.* Proc Natl Acad Sci U S A, 1980. **77**: p. 3225-3229.
7. Shinagawa, H., et al., *RecA protein-dependent cleavage of UmuD protein and SOS mutagenesis.* Proc Natl Acad Sci U S A, 1988. **85**: p. 1806-1810.
8. The NER repair is sometimes referred to as the "dark repair" mechanism. In the presence of light there is an alternative pathway using E. coli DNA photolyase which can fix cyclobutane pyrimidine dimers (Brash et. al., 1985, J Biol Chem, 260, 11438-11441.)
9. Setlow, R.B. and W. Carrier, *The disappearance of thymine dimers from DNA: an error correcting mechanism.* Proc. Natl. Acad. Sci. USA, 1964. **51**: p. 226-231.
10. Friedman, N., et al., *Precise Temporal Modulation in the Response of the SOS DNA Repair Network in Individual Bacteria.* PLoS Biol., 2005. **3**(7): p. e238.
11. Aksenov, S.V., *Dynamics of the inducing signal for the SOS regulatory system in Escherichia coli after ultraviolet irradiation.* Math Biosci, 1999. **157**: p. 269-286.
12. Sutton, M.D., T. Opperman, and G.C. Walker, *The Escherichia coli SOS mutagenesis proteins*

*UmuD and UmuD' interact physically with the replicative DNA polymerase.* Proc. Natl. Acad. Sci. USA, 1999. **96**: p. 12373-12378.
13. Cox, J.M., O.V. Tsodikov, and M.M. Cox, *Organized Unidirectional Waves of ATP Hydrolysis within a RecA Filament.* PLoS Biol., 2005. **3**: p. e52.
14. Battista, J.R., et al., *Dominant Negative UmuD Mutations Decreasing RecA-mediated Cleavage Suggest Roles for Intact UmuD in Modulation of SOS Mutagenesis.* Proc Natl Acad Sci U S A, 1990. **87**: p. 7190-7194.
15. Schnarr, M., et al., *DNA Binding Properties of the LexA Repressor.* Biochimie, 1991. **73**: p. 423-431.
16. McDonald, J.P., et al., *Intermolecular cleavage by UmuD-like mutagenesis proteins.* Proc Natl Acad Sci U S A, 1998. **95**: p. 1478-1483.
17. Frank, E.G., et al., *Regulation of SOS mutagenesis by proteolysis.* Proc Natl Acad Sci U S A, 1996. **93**: p. 10291-10296.
18. Rangarajan, S., R. Woodgate, and M.F. Goodman, *A phenotype for enigmatic DNA polymerase II: A pivotal role for pol II in replication restart in UV-irradiated Escherichia coli.* Proc. Natl. Acad. Sci. USA, 1999. **96**: p. 9224-9229.
19. Szekeres Jr., E.S., R. Woodgate, and C.W. Lawrence, *Substitution of mucAB or rumAB for umuDC Alters the Relative Frequencies of the Two Classes of Mutations Induced by a Site-Specific T-T Cyclobutane Dimer and the Efficiency of Translesion DNA Synthesis.* J. Bacteriol., 1996. **178**: p. 2559-2563.
20. Radman, M., *Enzymes of evolutionary change.* Nature, 1999. **401**: p. 866-869.
21. Opperman, T., et al., *A model for a umuDC-dependent prokaryotic DNA damage checkpoint.* Proc. Natl. Acad. Sci. USA, 1999. **96**: p. 9218-9223.
22. Yasuda, T., et al., *Physical interactions between DinI and RecA nucleoprotein filament for the regulation of SOS mutagenesis.* EMBO J, 2001. **20**: p. 1192-1202.
23. Ronen, M., et al., *Assigning numbers to the arrows: Parameterizing a gene regulation network by using accurate expression kinetics.* Proc Natl Acad Sci U S A, 2002. **99**: p. 10555-10560.
24. Lusetti, S.L., et al., *The DinI protein stabilizes RecA protein filaments.* J Biol Chem, 2004. **279**: p. 30037-30046.
25. Yoshimasu, M., et al., *An NMR study on the interaction of Escherichia coli DinI with RecA-ssDNA complexes.* Nucl. Acids Res., 2003. **31**: p. 1735-1743.
26. Voloshin, O.N., et al., *A model for the abrogation of the SOS response by an SOS protein: a negatively charged helix in DinI mimics DNA in its interaction with RecA.* Genes and Development, 2001. **15**: p. 415-427.
27. Crowley, D.J. and P.C. Hanawalt, *Induction of the SOS Response Increases the Efficiency of Global Nucleotide Excision Repair of Cyclobutane Pyrimidine Dimers, but Not 6-4 Photoproducts, in UV-Irradiated Escherichia coli.* J Bacteriol, 1998. **180**: p. 3345-3352.
28. Kelman, Z. and M. O'Donnell, *DNA Polymerase III Holoenzyme: Structure and Function of a Chromosomal Replicating Machine.* Annu. Rev. Biochem., 1995. **64**: p. 171-200.
29. Woodgate, R. and D.G. Ennis, *Levels of chromosomally encoded Umu proteins and requirements for in vivo UmuD cleavage.* Mol Gen Genet, 1991. **229**: p. 10-16.
30. Rupp, W.D. and P. Howard-Flanders, *Discontinuities in the DNA synthesized in an Excision-defective Strain of Escherichia coli following Ultraviolet Radiation.* J. Mol. Biol., 1968. **31**: p. 291-304.
31. Sommer, S., et al., *Specific RecA amino acid changes affect RecA-UmuD'C interaction.* Mol Microbiol, 1998. **28**: p. 281-291.